\documentclass[prd,twocolumn,showpacs,nofootinbib,amsmath,amssymb,floatfix]{revtex4}
\usepackage{graphicx}
\usepackage{times}
\usepackage{color}
\usepackage[a4paper,hyperindex=true]{hyperref}
\definecolor{linkcol}{rgb}{0,0,0.4} 
\definecolor{citecol}{rgb}{0.6,0,0} 

\hypersetup
{
bookmarksopen=true,
pdftitle="keV-DM",
pdfauthor="Hans HETTMANSPERGER", 
pdfsubject="keV sterile neutrino DM", %subject of the document
pdfmenubar=true, %menubar shown
pdfhighlight=/O, %effect of clicking on a link
colorlinks=true, %couleurs sur les liens hypertextes
pdfpagemode=UseOutlines, %aucun mode de page
pdfpagelayout=SinglePage, %ouverture en simple page
pdffitwindow=true, %pages ouvertes entierement dans toute la fenetre
linkcolor=linkcol, %couleur des liens hypertextes internes
citecolor=citecol, %couleur des liens pour les citations
urlcolor=linkcol %couleur des liens pour les url
}
\newcommand{\cL}{\mathcal{L}}
\newcommand{\diag}{\mathop\mathrm{diag}}
\newcommand{\eV}{\mbox{ \upshape\textrm{eV}}}
\newcommand{\keV}{\mbox{ \upshape\textrm{keV}}}
\newcommand{\TeV}{\mbox{ \upshape\textrm{TeV}}}
\newcommand{\MeV}{\mbox{ \upshape\textrm{MeV}}}
\newcommand{\GeV}{\mbox{ \upshape\textrm{GeV}}}

\newcommand{\abs}[1]{\mbox{$\left| #1 \right|$}}

\newcommand{\baz}{\begin{array}{cc}}
\newcommand{\ea}{\end{array}}
\newcommand{\be}{\begin{equation}}
\newcommand{\ee}{\end{equation}}
\newcommand{\Mpl}{M_\mathrm{Pl}}
\begin{document}
\title{keV sterile neutrino dark matter in gauge extensions
  of the standard model}

\author{F. Bezrukov}
\thanks{On leave from
  Institute for Nuclear Research of the Russian Academy of Sciences,
  60th October Anniversary prospect 7a, Moscow 117312, Russia.}
\email{Fedor.Bezrukov@mpi-hd.mpg.de}
\author{H. Hettmansperger}
\email{Hans.Hettmansperger@mpi-hd.mpg.de}
\author{M. Lindner}
\email{Manfred.Lindner@mpi-hd.mpg.de}
\affiliation{Max-Planck-Institut f\"ur Kernphysik,
  Postfach 103980, D-69029 Heidelberg, Germany}

\date{December 22, 2009}

%%%%%%%%%%%%%%%%%%%%%%%%%%%%%%%%%%%%%%%%%%%%%%%%%%%%%%%%%%%%%%%%%%%%%%%%
\begin{abstract}
  It is known that a keV scale sterile neutrino is a good warm dark
  matter candidate.  We study how this possibility could be
  realized in the context of gauge extensions of the standard model.
  The na\"ive expectation leads to large thermal overproduction
  of sterile neutrinos in this setup.  However, we find that it is
  possible to use out-of-equilibrium decay of the other right-handed
  neutrinos of the model to dilute the present density of the keV
  sterile neutrinos and achieve the observed dark matter density.  We present
  the universal requirements that should be satisfied by the gauge
  extensions of the standard model, containing right-handed neutrinos, to be
  viable models of warm dark matter, and provide a simple example in the context of
  the left-right symmetric model.
\end{abstract}
%%%%%%%%%%%%%%%%%%%%%%%%%%%%%%%%%%%%%%%%%%%%%%%%%%%%%%%%%%%%%%%%%%%%%%%%

\maketitle

%%%%%%%%%%%%%%%%%%%%%%%%%%%%%%%%%%%%%%%%%%%%%%%%%%%%%%%%%%%%%%%%%%%%%%%%
\section{Introduction}

Dark matter (DM) is one of the experimentally observed indications of
physics beyond the standard model (SM).  A wide variety of
astrophysical and cosmological observations confirm that
$\Omega_\mathrm{DM}\simeq0.2$ part of the total energy density of the
Universe is composed of some form of nonbaryonic matter which
interacts very weakly \cite{Bertone:2004pz}.
The most common particle physics explanation comes in the form of
weakly interacting massive particles, which are heavy and
weakly interacting thermal relics, leading to cold dark matter.
Another common candidate for cold dark matter is the axion, which is light, but
due to a specific generation mechanism it has an extremely small
temperature \cite{Turner:1989vc}.  Hot dark matter has high velocities
and a large free streaming length and it contradicts the experiment,
because it prevents the formation of the observed small scale
structures in the Universe.  The intermediate situation, warm dark
matter (WDM) is, however, less explored.  It may even provide a
solution to some of the problems of the DM simulations, reducing the
number of Dwarf satellite galaxies, or smoothing the cusps in the DM
halos.

A natural candidate commonly considered for WDM is a light sterile
neutrino \cite{Dolgov:2000ew}.\footnote{Note that WDM can also be many
  other particles, like a gravitino or even heavy particles; see
  \cite{Gorbunov:2008ka,Gorbunov:2008ui}.}  A simple realization is
the $\nu$MSM \cite{Asaka:2005an,Asaka:2005pn}, where only three
singlet fermions, which have Majorana masses and Dirac mixing with
ordinary (active) neutrinos, are added to the standard model.  Then,
the mass of one sterile neutrino can be chosen in the range of several
keV and with very small mixing with the active neutrinos, it will
provide a particle with the lifetime exceeding the age of the
Universe, which can be the WDM candidate.  The virtue and at the same
time the problem of the model is, that the sterile neutrino with such
a small mixing (the \emph{only} interaction of this particle is via
the Yukawa couplings) never enters into the thermal equilibrium, and
it can be produced only by some nonthermal mechanism.  If this were not
true and the neutrino reached thermal equilibrium at some moment in
the early universe, then without any additional mechanism, the thermal
relics with mass of over about 90~eV would overclose the Universe.  At the
same time, one needs knowledge of the physics before the beginning of
thermal evolution of the Universe in order to calculate unambiguously
the abundance of sterile neutrinos in the $\nu$MSM (see Refs.~\cite{Bezrukov:2008ut,Shaposhnikov:2006xi,Anisimov:2008qs}).

The possibility analyzed in this article is opposite to the $\nu$MSM.
We assume, that there is some additional (gauge) scale between the
electroweak and Planck scales, and that the sterile neutrinos are
charged under these additional gauge transformations.
% Leaving aside
% the naturalness of the small Yukawa couplings and low Majorana masses
% of the sterile neutrinos, we analyse the possibility to obtain WDM
% sterile neutrinos in such setup.

\begin{figure}
  \centering
  \includegraphics[width=\columnwidth]{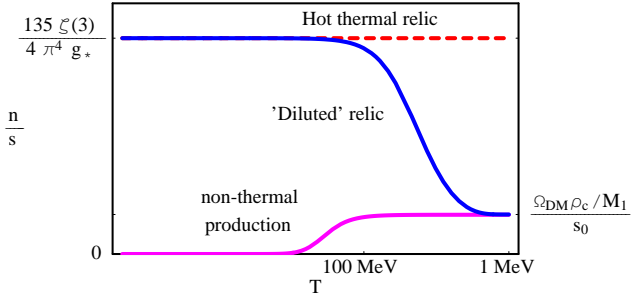}
  \caption{Schematic evolution of the light relic abundance in the
    Universe.  The dashed line is a thermal relic decoupled while
    being relativistic (hot thermal relic), leading to the overclosure
    of the Universe.  The blue decreasing line is the same hot thermal relic, but
    with the abundance diluted by rapid expansion of the Universe
    (entropy production), leading to correct DM abundance.  The lowest
    (magenta) line depicts the evolution of the nonthermally produced
    particle with zero primordial abundance.}
  \label{fig:nsratio}
\end{figure}

It turns out that it is possible to reconcile the thermal
overproduction of the DM with the observations.  To do this, the
abundance of the sterile neutrino should be diluted \emph{after} it
drops out of the thermal equilibrium.  This happens if some long-lived
particle decays while being out of thermal equilibrium after the DM
sterile neutrino freeze-out.  This effectively reduces the amount of
the DM sterile neutrino relative to the overall energy balance of the
Universe; see Fig.~\ref{fig:nsratio}.  It is also easy to find a
candidate for this long-lived heavy particle---another (heavier)
sterile neutrino in the model.  We formulate the requirements on the
properties of the DM sterile neutrino and the out-of-equilibrium
decaying particle to make the model consistent with existing
observations and bounds.  This generic analysis, important for all
possible models of this type, is made in the Sec.~\ref{sec:cosm-requ}.
In the end of this section, all the requirements are summarized.

There exist other ways to avoid the overproduction of the DM sterile
neutrino in the analyzed class of the models, which we will only
mention here.  One possibility is realised if all the new gauge
interactions are at the grand unified theory (GUT) scale, while the reheating after inflation
leads to temperatures below the GUT scale.  This situation is similar
to the $\nu$MSM, because the sterile neutrinos do not reach thermal
equilibrium.  Another possibility requires large (of the order of
thousand) number of degrees of freedom at the moment of the sterile
neutrino freeze-out, which does not seem natural.

In Secs.~\ref{sec:typeI} and \ref{sec:typeII}, we analyze the
possibility to realize these constraints in the simplest models. We then
use other sterile neutrinos to dilute the density of the DM
sterile neutrino.  In Sec.~\ref{sec:typeI}, we show that it is
impossible for the same right-handed neutrinos to be involved in the
DM abundance dilution and at the same time to give the masses to the
active neutrinos via a type I seesaw like mechanism.  The reason for
this is that the mixing angles (or, equivalently, Yukawa coupling
constants) are extremely small for the sterile neutrinos.  This would
lead to masses of the active neutrinos smaller, than the minimal ones,
allowed by the neutrino oscillation observations.

In Sec.~\ref{sec:typeII}, we provide a working example, where the
active neutrino masses are generated by a type II seesaw from the
scalar sector of a left-right (LR) symmetric model, and sterile
neutrinos have very small mixing angles with the active neutrino
sector.

The appendices are devoted to the calculation of the total decay width
of the sterile neutrinos in the model (Appendix \ref{appx:widths}),
radiative decay width (Appendix \ref{sec:raddecay}), and to the
description of the useful parametrization of the neutrino mass matrix
(Appendix \ref{CasasIbarra}).

%%%%%%%%%%%%%%%%%%%%%%%%%%%%%%%%%%%%%%%%%%%%%%%%%%%%%%%%%%%%%%%%%%%%%%%% 
\section{Cosmological requirements and constraints from experiments}
\label{sec:cosm-requ}

In this section, we introduce the generic framework we will work with
and discuss the various constraints and bounds resulting from
cosmological considerations and various experimental results.  Note
that these constraints are rather general and apply to most variations
of the specified model.

%%%%%%%%%%%%%%%%%%%%%%%%%%%%%%%%%%%%%%%%%%%%%%%%%%%%%%%%%%%%%%%%
\subsection{Assumptions and definitions}
\label{sec:model}

In the following, we will assume the existence of right-handed
(sterile) neutrinos $N_{IR}$.  These sterile neutrinos are not charged
under the SM gauge group, but they could be charged under the gauge
transformations of an extended model (ultimately, emerging in the
breaking chain of some GUT model).  Though for most of the statements
in this article the precise details of this gauge interaction are not
important, we will use a specific LR symmetric extension of the SM
and stick to it to obtain definite numbers.  This specific model (see
e.g.\ Ref.~\cite{Deshpande:1990ip} for a detailed review) with the gauge
group $SU(3)\times SU(2)_L\times SU(2)_R\times U(1)_{B-L}$ appears as
a subgroup of many GUT theories.

In this model, we have the interaction with the gauge bosons of the
form
\begin{multline}
  \label{eq:Lcc}
  -\cL_{CC} = \frac{g}{\sqrt{2}} \sum_a \left(
    W_L^\mu \,\overline{l_{aL}}\gamma_\mu\nu_{aL}
    +W_R^\mu \,\overline{l_{aR}}\gamma_\mu N_{aR}
  \right)\\
  +\textrm{h.c.}
  \;,
\end{multline}
where $W_L$ is the SM $W$ boson, $W_R$ is the corresponding right-handed boson 
from $SU(2)_R$, and $l_a$ are the charged SM leptons.
The neutrino mass matrix appears from the vacuum expectation values of
various Higgs bosons in the model.  Up to the Sec.~\ref{sec:typeII},
we will not be interested in the details of this, and will just write
the general mass matrix as
\begin{multline}
  \label{eq:Lmass}
  \cL_\textrm{mass} =
  -\frac12\left(\begin{array}{cc}
      \overline{\tilde\nu_{aL}^c}, \overline{\tilde N_{aR}}
    \end{array}\right)
  \left(
    \baz
    M_{L} & m_D \\
    m_D^T & M_{R}
    \ea
  \right)
  \left(\begin{array}{c}
      \tilde\nu_{aL}\\
      \tilde N_{aR}^c
    \end{array}\right)
  \\
  +\text{h.c.}
  \;.
\end{multline}
Note that a tilde over the neutrinos indicates that they are written
in the \emph{flavor} basis.  In the following, we will also assume
that the mass matrices obey in some sense the relations $M_R>m_D>M_L$
such that we can use seesaw-type formulas.  Thus, the rotation to the
mass basis has the form
\begin{multline}\label{eqn:trafo}
  \begin{pmatrix}
    \tilde\nu_{aL}\\
    \tilde N_{aR}^c
  \end{pmatrix} \simeq
  \\
  \begin{pmatrix}
                   1 & \left(M_{R}^{-1}m_D^T\right)^\dagger\\
    -M_{R}^{-1}m_D^T & 1
  \end{pmatrix}
  \begin{pmatrix}
    U & 0 \\
    0 & V_{R}
  \end{pmatrix}
  \begin{pmatrix}
    \nu_{iL}\\
    N_{IR}^c
  \end{pmatrix}
  \;,
\end{multline}
where $U$ is the standard Pontecorvo-Maki-Nakagawa-Sakata (PMNS) matrix and where $V_R$ describes the
mixing in the right-handed sector
\begin{align}
  \label{eq:see-saw}
  M_L-m_D M_R^{-1} m_D^T
  &= U^\star \cdot \diag(m_1,m_2,m_3) \cdot U^\dagger
  \;,\\
  M_R
  &= V_R^\star \cdot \diag(M_1,M_2,M_3) \cdot V_R^\dagger
  \;,
\end{align}
with $m_i$ being the active neutrino masses and $M_I$ the sterile
neutrino masses.  Note, that if $M_L=0$, then Eq.~(\ref{eq:see-saw}) is
the usual seesaw formula.

For the analysis of the sterile neutrino decay, when the oscillations
of the active neutrinos are not important, while the masses of the
charged leptons are, it is helpful to make the described rotation only
partially---without the PMNS rotation by the matrix $U$.  Then, we
get the mixing angles between the mass states of the sterile neutrinos
and SM flavors
\begin{equation}
  \theta_{a I} \equiv \frac{(m_D V_R)_{a I}}{M_I}
  \;,
\end{equation}
and also
\begin{equation}
  \label{eq:theta2def}
  \theta_I^2 \equiv \sum_{a=e,\mu,\tau} |\theta_{aI}|^2
  \;.
\end{equation}
These squared mixing angles describe the overall strength of
interaction (decay) of sterile neutrinos with the SM particles.

Before moving on to the analysis of the cosmological properties of
sterile neutrinos, let us note an additional possible complication.
Specifically, the $W_L$ and $W_R$ bosons in Eq.~(\ref{eq:Lcc}) may not
coincide with the mass eigenstates, $W_1$ and $W_2$ with masses $M_W$
and $M$, respectively, but be slightly mixed
\begin{align}\label{eq:Wmix}
  W_L &=  \cos\zeta\, W_1+\sin\zeta\, W_2\;,\notag \\
  W_R &= -\sin\zeta\, W_1+\cos\zeta\, W_2\;.
\end{align}
Normally this can be neglected, but it may give significant
contribution to the radiative decay of the DM sterile neutrinos,
analyzed in Sec.~\ref{sec:X-ray-obs}.

%%%%%%%%%%%%%%%%%%%%%%%%%%%%%%%%%%%%%%%%%%%%%%%%%%%%%%%%%%%%%%%%
\subsection{Temperature of freezeout}
\label{par:freezeout-temp}

Let us now calculate the moment of decoupling of the neutrinos $N_1$
in the early universe.  We will denote values corresponding to this
moment by the subscript ``f''.  As far as the DM sterile neutrino is
relatively light and the freeze-out happens while it is still
relativistic, the calculation is analogous to those for the usual
active neutrinos \cite{Kolb:1990vq}.  The only difference is that the
annihilation cross section is suppressed by the larger mass $M$ of the
right-handed gauge boson $W_R$, compared to the SM $W$ boson mass
$M_W$,
\begin{equation}
  \label{sigma-annihilate}
  \sigma_{N_1N_1}
  \approx \sigma_{\nu\bar\nu}\left(\frac{M_W}{M}\right)^4
  \sim G_F^2E^2\left(\frac{M_W}{M}\right)^4
  \;.
\end{equation}
Here, $\sigma_{\nu\bar\nu}$ is the SM neutrino annihilation cross
section, $G_F=1.166\times10^{-5}\GeV^{-2}$ is the Fermi constant and $E$ is
the energy of the colliding neutrinos.  Requiring the equality of the
mean free path and the Hubble scale, we get for the freeze-out
temperature
\begin{align}
  T_\mathrm{f}
  & \sim
  g_{*\mathrm{f}}^{1/6} \left(\frac{M}{M_W}\right)^{4/3} (1\div2)\MeV
  \;,
  \label{Tf}
\end{align}
where $g_{*\mathrm{f}}$ is the effective number of degrees of freedom
immediately after freeze-out (at least 10.75 for SM content if
freezeout happened below 100~MeV).

We see that for the not very large scale $M$, the sterile neutrino $N_1$
decouples at a rather low temperature.  Thus, it normally is in thermal
equilibrium at the early stages of the Universe evolution, making it a
thermal relic.  This will be the possibility which we peruse in the
current study.  In this case, calculation of the present day density of
the sterile neutrinos is insensitive to the history of the Universe
before $T_\mathrm{f}$.

Note, however, that if the reheating temperature after inflation is
lower than Eq.~(\ref{Tf}), the neutrinos never enter the thermal
equilibrium.  In this case, additional assumptions about the initial
abundance of the sterile neutrinos are necessary to predict their
current density, and the generation mechanism is very different form
the analyzed here (see, e.g.\
Refs.~\cite{Boyarsky:2009ix,Asaka:2006nq,Dodelson:1993je,Shi:1998km}). Such a situation can be realized for a very low reheating temperature (see e.g.\ \cite{Khalil:2008kp}), or naturally if the right-handed scale is the
GUT scale, $M\sim M_\mathrm{GUT}$, leading to $T_\mathrm{f}\sim
M_\mathrm{GUT}$, and the reheating after inflation reached slightly
lower temperatures. Another way to implement this situation is perused
in the $\nu$MSM \cite{Asaka:2005an,Asaka:2005pn}, where \emph{no new
  physics} is present up to Planck scale, leading to $N_1$ never
entering the thermal equilibrium.

%%%%%%%%%%%%%%%%%%%%%%%%%%%%%%%%%%%%%%%%%%%%%%%%%%%%%%%%%%%%%%%%
\subsection{Abundance of $N_1$ at present time}
\label{sec:N1-abundance}

The number to entropy density ratio of the sterile neutrino (two
fermionic degrees of freedom) after freeze-out is given by
\begin{equation}
  \left.\frac{n_{N_1}}{s}\right|_\mathrm{f} =
  \frac{1}{g_{*\mathrm{f}}}\frac{135\zeta(3)}{4\pi^4}
  \;. \label{eqn:DensityN1}
\end{equation}
While the Universe expands slowly with all the processes approaching
thermal equilibrium, both the number density and entropy density scale are
inversely proportional to the volume of the Universe, and this ratio
remains constant.  If nonequilibrium processes happen during
expansion (for example an intermediate matter dominated stage caused
by out of equilibrium decay of a heavy species), additional entropy
release is possible, which we will take into account by the factor $S$:
\begin{equation}
 \frac{n_{N_1}(t_0)}{n_{N_1}(t_\mathrm{f})}=
 \left(\frac{a(t_\mathrm{f})}{a(t_0)}\right)^3 =
 \frac{s(t_0)}{s(t_\mathrm{f})} \frac{1}{S}
 \;.
\end{equation}
Let us calculate the contribution of $N_1$ to the present energy
density.  Rescaling the number to entropy density ratio at present
moment by this factor, as compared to the freeze-out moment, we get for
the sterile neutrino contribution to the energy density of the
Universe $\Omega_N$
\begin{align}
  \frac{\Omega_N}{\Omega_\mathrm{DM}}
  &=
  \left(\left.\frac{n_{N_1}}{s}\right|_\mathrm{f}\right) \frac{1}{S} M_1
  \frac{s_0}{\Omega_\mathrm{DM}\rho_c}
  \notag\\
  &\simeq
  \frac{1}{S}\left(\frac{10.75}{g_{*\mathrm{f}}}\right)
  \left(\frac{M_1}{1\keV}\right)\times100
  \;,
  \label{OmegaDM}
\end{align}
where $\Omega_\mathrm{DM}=0.105h^{-2}$ is the DM density,
$s_0=2889.2$~cm$^{-3}$ is the present day entropy density, and
$\rho_c=1.05368\times10^{-5}h^2$~GeV~cm$^{-3}$ is the critical density of
the Universe.  The observational requirement is $ \Omega_N/\Omega_\mathrm{DM}
\leq 1$ with equality being the nicest choice (all DM is made out of
$N_1$) and inequality opting for multispecies DM.

Let us analyze Eq.~\eqref{OmegaDM} further.  Without entropy release
($S=1$), the Universe is overclosed, unless the neutrino is very light,
which corresponds to the hot dark matter case, excluded by the
structure formation in the Universe.  Models with the number of
degrees of freedom at freeze-out $g_{*\mathrm{f}}$ of order 1000 seem
rather unnatural and will not be considered.  The only opportunity is
thus the entropy release after freeze-out of $N_1$,
\begin{equation}
  \label{Sreq}
  S \simeq 100 \left(\frac{10.75}{g_{*\mathrm{f}}}\right)
  \left(\frac{M_1}{1\keV}\right)
  \;.
\end{equation}
Having this entropy release after $N_1$ decoupling will lead to the
observed DM abundance today.  In the following, we will analyze
possibilities of generation of this large amount in the
model.\footnote{The exact value of the required entropy release $S$ may be
  slightly different if, for example, some amount of DM sterile
  neutrino was generated nonthermally after the freezeout.  In the
  examples analyzed in the paper, this effect is negligible.}

%%%%%%%%%%%%%%%%%%%%%%%%%%%%%%%%%%%%%%%%%%%%%%%%%%%%%%%%%%%%%%%%
\subsection{Mass bounds}
\label{sec:MassBounds}

The mass of the DM particle can not be too light, or the observed
structure in the Universe would have been erased by a too hot DM.  The
simplest and most robust bound can be obtained from the phase space
density arguments.  The phase space density of a collisionless DM can
only become smaller during the evolution of the Universe, as an effect
of coarse-graining.  Comparison of the primordial phase space density,
which is calculated using the initial DM particle distribution
function and the maximal modern one, derived from the observation of
the Dwarf spheroidal galaxies \cite{Gorbunov:2008ka,Boyarsky:2008ju}
gives the lower bound
\begin{equation}
  M_1 > 1\text{--}2\ \mathrm{keV}
  \;.
\end{equation}

Another important bound is the Lyman-$\alpha$ (Ly-$\alpha$) bound
\cite{Boyarsky:2008xj,Seljak:2006qw}.  This bound constrains the
velocity distribution of the DM particles from the effect of their
free streaming on the formation of the structure on the scales, probed
by the Ly-$\alpha$ forest.  It should be noted that to convert this
constraint into a bound for the mass of the DM particle, one needs to
take into account the initial velocity distribution of the particles.
In our case it takes the form of a usual thermal distribution, but
with the temperature lowered by the dilution factor $S^{-1/3}$.  This
corresponds to the \emph{thermal relic} case in
Ref.~\cite{Boyarsky:2008xj}, and not to the case of the nonresonantly
produced sterile neutrinos, denoted $m_\mathrm{NRP}$ in
Ref.~\cite{Boyarsky:2008xj}.  Thus, the result of Ref.~\cite{Boyarsky:2008xj}
should be rescaled as
\begin{equation}
  \label{LyaS}
  M_1 > \frac{T}{T_\nu}m_\mathrm{NRP} \simeq 1.6\ \mathrm{keV}
  \;,
\end{equation}
where $T$ is the present temperature of the DM neutrino diluted with
the entropy factor (\ref{Sreq}), $T_\nu$ is the temperature of the
usual relic neutrinos, $m_\mathrm{NRP}=8$~keV
\cite{Boyarsky:2008xj}, and the ratio of the temperatures
$(T/T_\nu)^3=\Omega_\mathrm{DM}h^2(94\text{ eV}/M_1)$ is obtained from
the requirement of the observed $\Omega_\mathrm{DM}$.

%%%%%%%%%%%%%%%%%%%%%%%%%%%%%%%%%%%%%%%%%%%%%%%%%%%%%%%%%%%%%%%%
\subsection{Generation of entropy }
\label{sec:GenerationEntropy}

The entropy \eqref{Sreq} can be generated by some heavy long-lived
particle which goes out of the thermal equilibrium at some moment
after DM neutrino freeze-out $t_\mathrm{f}$, and it decays after becoming
nonrelativistic and dominating the Universe expansion.  The obvious
candidates for such particles are the two remaining heavier neutrinos
(though other candidates are possible and can be analyzed in a
similar way).  Let us assume for simplicity that only one of these two
neutrinos is responsible for entropy generation, and we denote it by
$N_2$.  Then, according to Refs.~\cite{Scherrer:1984fd,Kolb:1990vq}, the
entropy release is
\begin{equation}
  \label{Sgeneric}
  S\simeq \left(
    1 + 2.95 \left(\frac{2\pi^2\bar{g}_{*}}{45}\right)^{1/3}
        \frac{(rM_2)^{4/3}}{(\Mpl\Gamma)^{2/3}}
  \right)^{3/4}
  \;,
\end{equation}
where $M_2$ is the mass of $N_2$, $r=n_{N_2}/s$ is the initial
abundance of the $N_2$ particles after decoupling (or, probably more
precise, before they start to drive the matter dominated intermediate
stage of the Universe expansion), and $\bar{g}_{*}$ is the properly
averaged effective number of degrees of freedom during the $N_2$ decay.  The
ratio $r$ is maximal when the particle decouples when it is still
relativistic, and is equal to
\begin{equation}
  r\equiv\frac{n_{N_2}}{s}=\frac{g_{N}}{2}\frac{135\zeta(3)}{4\pi^4g_{*}}
  \;,
  \label{eqn:r}
\end{equation}
where $g_{N}=2$ is the number of degrees of freedom for $N_2$, and $g_{*}$ is
taken at the $N_2$ freeze-out.

If the entropy generation is large, we can neglect the $1$ in
Eq.~(\ref{Sgeneric}) and get
\begin{equation}\label{eqn:SDecayWidth}
  S
  \simeq
  0.76\frac{g_{N}}{2}\frac{\bar{g}_{*}^{1/4}M_2}{g_{*}\sqrt{\Gamma \Mpl}}
  \;.
\end{equation}
By combining Eqs.~\eqref{Sreq} and \eqref{eqn:SDecayWidth}, we obtain
\begin{equation}\label{GammaSgen}
  \Gamma
  \simeq
  0.50\times10^{-6}
  \frac{g_N^2}{4} \frac{g_{*\mathrm{f}}^2}{g_{*}^2}
  \bar{g}_{*}^{1/2}\frac{M_2^2}{\Mpl}
  \left(\frac{1\keV}{M_1}\right)^2
  \;.
\end{equation}
Note that for our case, the freeze-out temperatures of the DM sterile
neutrino and of the entropy generating one coincide, so
$g_*=g_{*\mathrm{f}}$.  If Eq.~(\ref{GammaSgen}) is satisfied, then we
have proper DM abundance in the present Universe.

However, Eq.~(\ref{GammaSgen}) is not the only requirement for the
lifetime of the heavier sterile neutrino for the realistic model.
Entropy generation should finish before the big bang nucleosynthesis
(BBN), i.e.\ $N_2$ should decay before it.  According to
Refs.~\cite{Asaka:2006ek,Kawasaki:2000en,Hannestad:2004px}, the temperature
after the decay of the sterile neutrino $N_2$ should be greater than
$0.7\div4\,\MeV$ in order not to spoil BBN predictions.  This
temperature is approximately equal to (see
Ref.~\cite{Scherrer:1984fd})
\begin{equation}
  T_r \simeq
  \frac{1}{2}\left(\frac{2\pi^2\bar{g}_{*}}{45}\right)^{-1/4}
  \sqrt{\Gamma \Mpl}
  \;,
\end{equation}
leading to the bound on the $N_2$ lifetime which should be shorter
than approximately $0.1\div2$~s.  The neutrino with such a lifetime
can produce enough entropy, satisfying Eq.~(\ref{GammaSgen}) only if
it is sufficiently heavy,
\begin{equation}
  \label{BBNbound}
  M_2 > \left(\frac{M_1}{1\keV}\right) (1.7\div10)\GeV
  \;.
\end{equation}
Finally, as far as we were assuming that the sterile neutrino $N_2$
decoupled while still relativistic (otherwise the entropy generation
is much less efficient), we should require $T_\mathrm{f}>M_2$.  This,
using Eq.~(\ref{Tf}), is translated into a bound for the scale of the
right-handed bosons,
\begin{equation}\label{BOUNDMR}
  M >
  \frac{1}{g_{*\mathrm{f}}^{1/8}}\left(\frac{M_2}{\GeV}\right)^{3/4}(10\div16)\TeV
  \;.
\end{equation}
Thus, on the one hand the sufficient entropy generation requires a
long-lived neutrino, but on the other hand, the requirement of the
successful BBN limits its lifetime from above, leading to the lower
bounds on its mass and on the mass scale of the additional gauge
interactions.

%%%%%%%%%%%%%%%%%%%%%%%%%%%%%%%%%%%%%%%%%%%%%%%%%%%%%%%%%%%%%%
\subsection{x-ray observations}
\label{sec:X-ray-obs}

\begin{figure}
  \includegraphics[width=0.9\columnwidth]{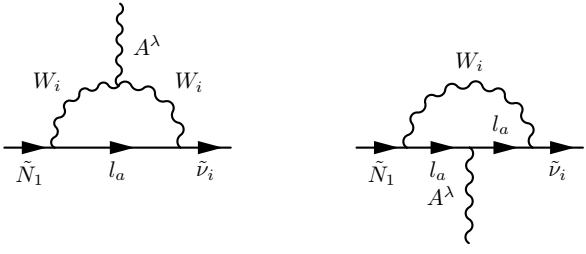}
  \caption{\label{fig:DecayN}Unitary-gauge diagrams contributing to
    the radiative neutrino decay with charged leptons propagating in
    the loop.}
\end{figure}

A sterile neutrino in the considered class of the models is unstable,
so it provides a \emph{decaying} DM.  Through its mixing, it decays via
the neutral current into three active neutrinos.  To lead to a
successful DM scenario, the lifetime of the unstable neutrino $N_1$
should be greater than the age of the Universe
$\tau_\mathrm{u}\sim10^{17}\sec$, which constrains its total decay
width.  However, one obtains significantly stronger restrictions
resulting from a subdominant decay channel---the radiative decay
$N_1\to\nu\gamma$, induced at the one loop level
(Fig.~\ref{fig:DecayN}).  This process produces a narrow line in the
x-ray spectrum of astrophysical objects
\cite{Dolgov:2000ew,Abazajian:2001vt}.  In the context of $\nu$MSM, the
only source of this decay is via the active-sterile neutrino mixing
$\theta_1^2$, and the recent x-ray observations bound it from above.
A very rough bound, which will be enough for our purposes, is given in
Ref.~\cite{Boyarsky:2009ix}\footnote{We must note that careful analysis
  gives a stronger (in some regions of masses by an order of
  magnitude) bound.  See the detailed discussion in Sec.~5.1.2 of
  \cite{Boyarsky:2009ix} and
  \cite{Boyarsky:2006fg,Boyarsky:2006ag,Boyarsky:2007ay,Boyarsky:2006hr,Boyarsky:2007ge,Watson:2006qb,Yuksel:2007xh}.  For our
  purposes, this approximate (weak) bound is sufficient.}
\begin{equation}\label{theta-Xraybound}
  \theta_1^2\lesssim1.8\times10^{-5}\left(\frac{1\keV}{M_1}\right)^5
  \;.
\end{equation}
This bound corresponds to the following upper bound on the radiative
decay width
\begin{equation}\label{eqn:Widthbound}
  \Gamma_{N_1\to\gamma\nu}\lesssim9.9\times10^{-27}\sec^{-1}
  \;.
\end{equation}

We also note that there are bounds resulting from supernova cooling.
They are also much weaker than the diffuse x-ray background limits
\eqref{theta-Xraybound} for all possible neutrino masses $M_1$
\cite{Dolgov:2000ew}.

In the LR symmetric model, the x-ray bound (\ref{eqn:Widthbound}) leads
not only to the bound on the mixing angle (\ref{theta-Xraybound}), but
also bounds the properties of the bosonic sector of the theory.  The
reason is the possible mixing of the right $W_R$ gauge bosons with the
SM $W_L$ ones.  Without mixing, the contribution of the $W_R$ bosons
to the process $N\to\gamma\nu$ is additionally suppressed by the ratio
of the left and right gauge boson masses $(M_W/M)^4$, and can be
safely neglected.  With the mixing, however, the chiral structure of
the diagram changes, and the contribution can be enhanced by the
factor $(m_l/M_1)^2$, where $m_l$ is the mass of the charged lepton
running in the loop.

We calculate the total decay width for $N_1\to\gamma\nu$, summed over
the active neutrino flavors, following
Refs.~\cite{Lavoura:2003xp,Denner:1992me} (for details, see Appendix \ref{sec:raddecay}).  Supposing from the very beginning that the
right-handed scale is much larger than the left one, $M\gg M_W$,
neglecting the active neutrino masses and assuming small gauge boson
mixing, we get
\begin{equation}\label{eqn:zetanonzero}
  \Gamma_{N_1\rightarrow\gamma\nu}
  \simeq \frac{G_F^2\alpha M_1^3}{64\pi^4}
  \sum_{a=e,\mu,\tau}
  \left| 4 m_{l_a} (V_R)_{a1}\cdot \zeta
    -\frac{3}{2} \theta_{a1} M_1 \right|^2
  \;.
\end{equation}
Here, $\alpha$ is the fine-structure constant, and $m_{l_a}$ is the
mass of the charged lepton propagating in the loop.  The second term
in the amplitude is proportional to the mass of the sterile neutrino
$M_1$, while the first term to the mass of the charged intermediate
lepton $m_{l_a}$.  This can be understood from the following
consideration.  Because the photon has spin one, there must be a
chirality flip on the fermionic line.  If, in flavor basis, there is
a $W_R$-$W_L$ mixing, we have the chirality flip (mass insertion) on the
internal line of the charged fermion, which produces a term
proportional to $m_{l_a}$.  Otherwise, the chirality flip happens on
one of the outer lines of the diagram, with a term proportional to the
Majorana mass of the incoming sterile neutrino.

If the gauge boson mixing is absent, $\zeta=0$, only the second term
contributes and we obtain the usual result
\begin{equation}\label{eqn:zetazero}
  \Gamma_{N_1\rightarrow\gamma\nu}\Big|_{\zeta=0}
  \simeq \frac{9\,G_F^2\,
    \alpha\,M_1^5}{256\, \pi^4}\times\theta_1^2
  \;.
\end{equation}
If the mixing $\zeta$ is present, then, barring the unlikely
cancellation between the two terms in Eq.~(\ref{eqn:zetanonzero}), we can
constrain $\zeta$ using the x-ray bound (\ref{eqn:Widthbound})
\begin{equation}\label{zetabound}
  \zeta^2 \lesssim 9\times10^{-19}\,
  \frac{m_{l_\tau}^2}{\sum_{a=e,\mu,\tau}\abs{m_{l_a}(V_R)_{a1}}^2}
  \,\left(\frac{\keV}{M_1}\right)^3
  \;.
\end{equation}
Thus, the mixing angle of the $W$ bosons must be vanishingly small.

We would like to note that in a LR symmetric model with the Higgs
sector as described in Sec.~\ref{sec:typeII}, the $W_L$-$W_R$ mixing
angle $\zeta$ is given by (see Ref.~\cite{Deshpande:1990ip}, and references
therein)
\begin{equation}\label{eqn:tan(2zeta)}
  \tan(2\zeta)\simeq-\frac{2\kappa_1\kappa_2}{v_R^2}
  \,,
\end{equation}
where $\kappa_1$ and $\kappa_2$ are bidoublet vacuum expectation values (VEVs) and $v_R$ is the
VEV of the right Higgs triplet. Therefore Eq.~\eqref{zetabound}
restricts the ratio $\kappa_1\kappa_2/v_R^2$.

%%%%%%%%%%%%%%%%%%%%%%%%%%%%%%%%%%%%%%%%%%%%%%%%%%%%%%%%%%%%%%%%
\subsection{Summary of constraints}
\label{sec:summary}

Let us summarize this section.  A theory where the DM sterile neutrino
was in thermal equilibrium at some moment during the evolution of the
universe should satisfy the following set of constraints:

\begin{itemize}
\item From $X/\gamma$-ray observations, we have the model independent
  upper limit on the radiative decay width of the DM neutrino $N_1$
  (see Eq.~\eqref{eqn:Widthbound}):
  \begin{equation}
    \Gamma_{N\to\gamma\nu}\lesssim9.9\times10^{-27}\sec^{-1}
    \;.
  \end{equation}
  Note that this is a conservative value; c.f.\ footnote on page
  \pageref{theta-Xraybound}.

  This limit translates to the limit on the sterile-active neutrino
  mixing angle Eq.~(\ref{theta-Xraybound}) and to the bound on the
  mixing between the left and right gauge bosons
  Eq.~(\ref{zetabound}).
\item From the structure formation requirements (Ly-$\alpha$ bound),
  the mass of the sterile neutrino is constrained in the same way as
  the mass of a thermal relic, i.e.
  \begin{equation}
    M_1\gtrsim1.6\keV
    \;.
  \end{equation}
\item The right abundance of the sterile neutrino can be then achieved
  by an out-of-equilibrium decay of a long-lived heavy particle.  We
  will use another sterile neutrino of the model, $N_2$, for this
  purpose, but most considerations here can be also applied to another
  long-lived particle present in the theory.

  To provide proper the entropy dilution, Eq.~(\ref{Sreq}), $N_2$
  should decouple while relativistic and has decay width
  \begin{equation}
    \Gamma \simeq
    0.50\times10^{-6}
    \frac{g_N^2}{4} \frac{g_{*\mathrm{f}}^2}{g_{*}^2}
    \bar{g}_{*}^{1/2}\frac{M_2^2}{\Mpl}
    \left(\frac{1\keV}{M_1}\right)^2
    \;.
  \end{equation}
\item At the same time, the heavy neutrino $N_2$ should decay before
  BBN, which bounds its lifetime to be shorter than approximately
  $0.1\div2$~s.  Then, the proper entropy can be generated only if
  its mass is larger than
  \begin{equation}
    M_2 > \left(\frac{M_1}{1\keV}\right)
    (1.7\div10)\GeV
    \;.
  \end{equation}
\item The entropy is effectively generated by out-of-equilibrium decay
  (see Sec.~\ref{sec:GenerationEntropy}), if the particle decoupled while still
  relativistic.  If this particle is one of the sterile neutrinos,
  then its decoupling happens at temperature (\ref{Tf}), and it leads
  to the bound on the right-handed gauge boson mass
  \begin{equation}
    M >
    \frac{1}{g_{*\mathrm{f}}^{1/8}}\left(\frac{M_2}{1\GeV}\right)^{3/4}(10\div16)
    \TeV
    \;.
  \end{equation}
  Note that this is the only requirement which changes in the case of
  entropy generated by some other particle instead of the heavy
  sterile neutrino.
\end{itemize}

%%%%%%%%%%%%%%%%%%%%%%%%%%%%%%%%%%%%%%%%%%%%%%%%%%%%%%%%%%%%%%%%%%%%%%%%
\section{Models with low scale type I see-saw}
\label{sec:typeI}

Let us start from the analysis of the models where the active
neutrino masses are generated by a ``type I'' seesaw
formula.  This means that $M_L=0$ in the neutrino mass matrix (\ref{eq:Lmass}).  The mixing angles
(\ref{eq:theta2def}) are bounded from above by the requirements on the
decay width of the sterile neutrinos---by the x-ray observations for the DM neutrino angle $\theta_1$ (see Eq.~(\ref{theta-Xraybound})), and by
the long enough lifetime of the entropy generating neutrino angle
$\theta_2$ (additional generation of the entropy by the third neutrino
does not change conclusions).  A convenient way to parameterise the
Dirac mass matrix $m_D$, separating parameters in the active and
sterile neutrino sectors, is provided by the Casas and Ibarra
parametrization \cite{Casas:2001sr} reviewed in
Appendix \ref{CasasIbarra}.  Using Eq.~\eqref{eqn:CasasIbarra} for the Dirac masses,\footnote{As far as we are
  using in this section the basis with diagonal $M_R$, the right-handed mixing matrix is trivial, $V_R = I$.} we get
\begin{equation}\label{eqn:Theta}
\theta_I^2=\frac{\left[\sqrt{M_R}\,
R^T\,m_\nu^{\diag}\,R^\star\,\sqrt{M_R}\right]_{II}}{M_I^2 }
\;,
\end{equation}
with
\begin{equation}
m_\nu^{\diag} = \diag(m_1,m_2,m_3)
\;.
\end{equation}
Here, $R$ is a complex orthogonal matrix, describing the details of the mixing between the sterile and active sectors, and it can be parameterized by three complex angles $\omega_{12}$, $\omega_{13}$, and $\omega_{23}$ as in Eq.~(\ref{eqn:R}).
Let us check whether we can satisfy the bounds on the mixing angles if the active masses $m_i$ are consistent with the observed neutrino oscillation mass differences, summarized below. The current best-fit and 3$\sigma$ ranges are (see Ref.~\cite{Schwetz:2008er})
\begin{subequations}
\begin{align}
\Delta m_{\mathrm{sol}}^2 = \left(7.65^{+0.69}_{-0.6}\right) \cdot 10^{-5} \eV^2\;,\\
\Delta m_{\mathrm{atm}}^2 = \left(2.4^{+0.35}_{-0.33}\right) \cdot 10^{-3} \eV^2\;.
\end{align}
\end{subequations}
In the following discussion we will for convenience order the active neutrino masses as $m_1<m_2<m_3$. From Eq.~(\ref{eqn:Theta}), we get for the first two sterile neutrinos
\begin{subequations}\label{eqn:M1thetaM2theta}
\begin{align}
M_1\theta_1^2 &= m_3\,\abs{\sin{\omega_{13}}}^2 + m_2\abs{\cos{\omega_{13}}}^2\abs{\sin{
\omega_{12}}}^2\notag\\&\qquad +m_1\abs{\cos{\omega_{13}}}^2\abs{\cos{
\omega_{12}}}^2\;,\\
M_2\theta_2^2 &= m_3\abs{\cos{\omega_{13}}}^2\abs{\sin{\omega_{23}}}^2\notag\\
&\qquad+m_2\abs{\cos{\omega_{23}}\cos{\omega_{12}}-\sin{\omega_{23}}\sin{\omega_{13}}\sin
{
\omega_{12}}}^2\notag\\&
\qquad+m_1\abs{\cos{\omega_{23}}\sin{\omega_{12}}+\sin{\omega_{23}}\sin
{\omega_{13}}\cos{\omega_{12}}}^2
\;.\label{eqn:B*}
\end{align}
\end{subequations}
Note that as far as we ordered the active neutrino masses, if we change $m_1$ to zero, and replace $m_3$ by $m_2$, the right-hand sides of Eqs. in \eqref{eqn:M1thetaM2theta} can only become smaller.  We can also confine ourselves to the real values of the mixing angles, as far as the sine and cosine absolute values only become larger for complex angles, and the inequality $\abs{z-w}\geq\abs{\abs{z}-\abs{w}}$ is used to transform the square of the difference of the angles in Eq.~(\ref{eqn:B*}).  Thus, the following inequalities should be satisfied:
\begin{subequations} \label{eqn:M1M2thetaInequ}
\begin{align}
M_1\theta_1^2 &\geq m_2\{\sin^2{\omega_{13}} +\cos^2{\omega_{13}}\sin^2{\omega_{12}}\}\;,\label{eqn:M1theta}\\
M_2\theta_2^2 &\geq m_2\Big\{\cos^2{\omega_{13}}\sin^2{\omega_{23}}+\big(\abs{\cos{\omega_{23}}}\abs{\cos{\omega_{12}}}\notag\\
&\qquad\qquad-\abs{\sin{\omega_{23}}}\abs{\sin{\omega_{13}}}\abs{\sin{
\omega_{12}}}\big)^2\Big\}
\;.\label{eqn:M2theta}
\end{align}
\end{subequations}
The minimum of the sum of the right-hand sides is $m_2$, and therefore the following very simple inequality always holds:
\begin{equation}\label{eqn:boundM1M2theta}
  M_1\theta_1^2+M_2\theta_2^2 \ge m_2 \ge\Delta m_\mathrm{sol}
  \;.
\end{equation}
The second inequality is trivially fulfilled, since in all possible mass hierarchies the mass of the second (in mass) active neutrino is larger than $\Delta m_\mathrm{sol}$.  The meaning of the inequality (\ref{eqn:boundM1M2theta}) is very simple---one can not generate active neutrino masses with type I seesaw formula without sufficient mixings between the active and sterile neutrino sectors. Note in passing that the cancellation is possible in another direction---one can have very small active neutrino masses and large active-sterile mixings.

Now, we are ready to compare the requirement from the observed active neutrino masses, Eq.~(\ref{eqn:boundM1M2theta}), and the DM bounds on the mixings.
The angle $\theta_2$ can be bound from the width required to generate sufficient entropy, Eq.~(\ref{GammaSgen}).  Estimating the width of the heavy neutrino as (see Appendix \ref{appx:widths})
\begin{align}
\Gamma_{N_2} 
	&\geq \frac{G_F^2M_2^5}{192\pi^3}\cdot\theta_2^2
\;,\label{eqn:GammaLowerBound}
\end{align}
we have
\begin{align}
M_2\theta_2^2&\lesssim
1.8\times10^{-3} \bar{g}_*^{1/2} \left(\frac{\GeV}
{M_2}\right)^2\left(\frac{\keV}{M_1}\right)^2
\;.
\label{eqn:theta_2^2}
\end{align}
It can be clearly seen that for all possible masses $M_1$ and $M_2$, this
is much smaller than $\Delta m_\mathrm{sol}$.

The contribution of the DM sterile neutrino itself can be larger.
From Eq.~(\ref{theta-Xraybound}), we have
\begin{equation}\label{eq:M1th1}
  M_1\theta_1^2 \lesssim
  1.8\times10^{-2}\left(\frac{1\keV}{M_1}\right)^4
  \;.
\end{equation}
Together with the Ly-$\alpha$ bound on the WDM mass, Eq.~(\ref{LyaS}), this contribution again violates Eq.~(\ref{eqn:boundM1M2theta}).  Thus, we conclude that the small mixing angles, required by the proper DM abundance and good DM properties in the model, prevent generation of the observed active neutrino masses by the type I seesaw formula.\footnote{Note, however, that without the Ly-$\alpha$ bound it would have been possible for very light WDM, with $M_1<1.2\keV$.}

%%%%%%%%%%%%%%%%%%%%%%%%%%%%%%%%%%%%%%%%%%%%%%%%%%%%%%%%%%%%%%%%%%%%%%%%
\section{Type II seesaw---working example}
\label{sec:typeII}

In the previous section, we have seen that if one of the not DM-like
sterile neutrinos is responsible for entropy production, it is impossible to
get the observed active neutrino masses with the type I like seesaw.
Here, we will present a working example of the sterile neutrino DM in
the framework of a LR symmetric model, where the active
neutrino masses are generated by the contribution of the type II
seesaw.

We will continue to work in the framework of the $SU(3)\times
SU(2)_L\times SU(2)_R\times U(1)_{B-L}$ model, sketched in Sec.~\ref{sec:model}.  Here, we will concentrate on a properly LR symmetric model, where the left- and
right-handed leptons are treated symmetrically.  One has the usual SM
doublets $\psi_L^i;\; i=1,2,3,$ and in addition 3 right-handed
neutrinos which form together with the 3 charged right-handed leptons
the $SU(2)_R$ doublets $\psi_R^i$.  The Higgs sector consists of one
$SU(2)_L$ triplet, one $SU(2)_R$ triplet, and one bidoublet.  In such a
model, the mass matrix for the neutrinos has the pattern
\begin{equation}
  \label{eqn:see-saw12}
  {\cal M} = \left( \baz
    f_L\,v_L & \ y\,v \\
    y^T\,v & f_R \,v_R \ea \right)=\left( \baz
    M_{L} & m_D \\
    m_D^T & M_{R} \ea \right)
  \;,
\end{equation}
where the Majorana blocks on the diagonal come from the coupling of
${\psi^i_L}^ T{\cal C} {\psi^j_L}$ and ${\psi^i_R}^T{\cal C}
{\psi^j_R}$ with the triplets $\Delta_{L,R}$, respectively, and the
Dirac-type ones from the coupling of $\bar\psi^i_L\psi^j_R$ with
the bidoublet $\phi$ and its complex conjugate $\tilde
\phi=\tau_2\phi^\star\tau_2$. The VEVs of the neutral components in
$\Delta_{L,R}$ are called $v_{L,R}$; whereas, the SM scale
$v=\sqrt{\kappa_1^2+\kappa_2^2}=174 \GeV$ is a combination of the
bidoublet VEVs $\kappa_1$ and $\kappa_2$.  These VEVs are related by
the expression
\begin{equation}
  \label{eqn:x}
  x\equiv\frac{v_Lv_R}{v^2}
  \;,
\end{equation}
where $x$ is a function of the parameters in the Higgs potential,
which is naturally of order one (for more details, see e.g. Ref.~\cite{Deshpande:1990ip}).

In the following, we postulate \emph{exact} discrete LR
symmetry.  In general, it can be realized in two different ways: as a
$\cal{C}$ conjugation or as a parity symmetry.  In the former case, it is
required that $f_L = f_R$, $y = y^T$, and this is what we use.

With such a model, it is now possible to satisfy all the requirements
from Sec.~\ref{sec:cosm-requ}.  Let us consider the type II seesaw
formula following from block diagonalization of Eq.~\eqref{eqn:see-saw12}
and the assumption
$\mathcal{O}(M_{R})\gg\mathcal{O}(m_D)\gg\mathcal{O}(M_{L})$:
\begin{equation}
  m_\nu = v_Lf_L-\frac{v^2}{v_R}yf_R^{-1}y^T
  \;.\label{eqn:typeIIsee-saw}
\end{equation}
After applying the conditions of
discrete left-right symmetry,
 $y=y^T$ and $f\equiv f_L=f_R$, one
arrives at
\be
  m_\nu = v_Lf-\frac{v^2}{v_R}yf^{-1}y
  \;,\label{eqn:typeIIsee-saw*}
\ee
To simplify the calculations, we
further assume for illustration that the Dirac-Yukawa $y$ is
proportional to the triplet Yukawa $f$, i.e.\ $y=p\,f$, where $p$ is a
number.  Equation~\eqref{eqn:typeIIsee-saw*} then goes into
\be
  m_\nu = \left(v_L-\frac{v^2p^2}{v_R}\right)f
  \;.\label{eqn:typeIIsee-sawmod}
\ee
In this case, all Yukawas are diagonalized by the same
transformation---the transformation which brings $m_\nu$ into diagonal form, i.e.\ the
PMNS matrix. The ratios of the eigenvalues of the matrices on
both sides of the equality are then the same
\begin{equation}
  \frac{m_1}{m_2}=\frac{f_1}{f_2}=\frac{M_1}{M_2}
  \;.\label{eqn:m_1}
\end{equation}
Thus, the mass spectrum of the sterile neutrinos (or, specifically,
the BBN requirement \eqref{BBNbound}) leads to the same hierarchical
active neutrino spectrum
\begin{equation}
 \frac{m_1}{m_2}\lesssim 5.9\times10^{-7}
\;.
\end{equation}
This implies that the lightest active neutrino should be
very light, and we can have either normal or inverse hierarchy.  For
definiteness, we will use the normal hierarchy for our example, though
the inverse one works equally well (one should only take into account
that in the latter case the $M_2\simeq M_3$, $\Gamma_2\simeq\Gamma_3$
and both $N_1$ and $N_2$ generate the same amount of entropy). As far as the active neutrino mass
hierarchy is fixed, we have $m_2\simeq\Delta m_\mathrm{sol}$ and
$m_3\simeq\Delta m_\mathrm{atm}$.  We can then get the mass for the
third sterile neutrino from
\be\label{eqn:M_3}
M_3=\frac{m_3}{m_2}\,M_2
\;.
\ee
The active-sterile mixing angles in the case of proportional Yukawa
constants are all the same and equal to
\be\label{eqn:theta_2^2typeII}
\theta_1^2=\theta_2^2=\theta_3^2=\frac{v^2p^2}{v_R^2}
\;,
\ee
while the mixing angles for individual flavors are proportional to
the PMNS matrix
\begin{equation}
  \theta_{aI} = (U^\star)_{aI} \frac{vp}{v_R}
  \;.\label{eqn:thetaSecTypeII}
\end{equation}
Thus the decay width $\Gamma_2$ is proportional to $\theta_2^2$ (see
Appendix \ref{appx:widths}).  The value of $\theta_2^2$ is then
defined from the requirement of the sufficient entropy production,
Eq.~(\ref{GammaSgen}), and depends only on $M_1$ and $M_2$.

\begin{figure}
  \begin{flushright}
 \includegraphics{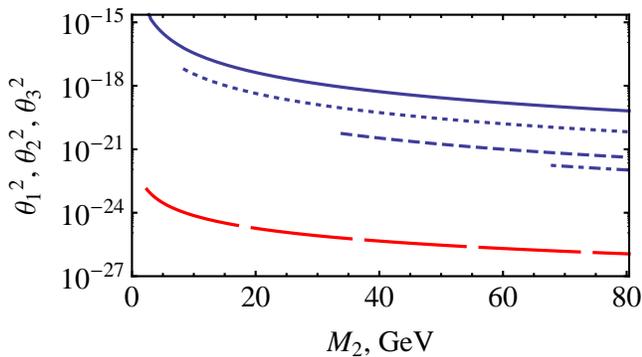}
  \end{flushright}
\caption{\label{fig:theta_2-M2}$\theta_2^2$ as a function of $M_2$. $M_1=1.6\keV$ (continuous); $M_1=5\keV$ (dotted); $M_1=20\keV$ (dashed); $M_1=40\keV$ (dashed-dotted). It is accounted for the lower bound, Eq.~\eqref{BBNbound}. The long-dashed (red) line shows the ratio $(M_{W}/M)^4$ and illustrates that processes mediated by $W_R$ bosons can be neglected in the decay rate of $N_2$.}
\end{figure}

\begin{figure}
  \begin{flushright}
    \includegraphics[scale=0.9]{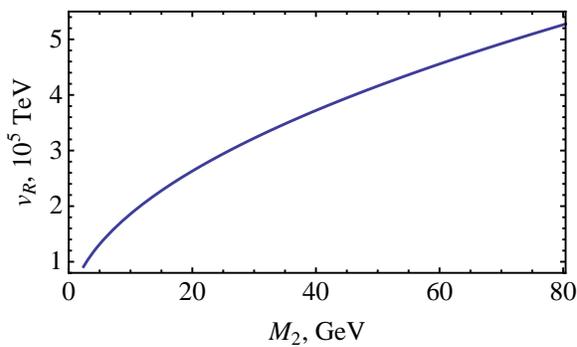}
  \end{flushright}
  \caption{\label{fig:M2-vR}$v_R$ as a function of $M_2$.  The dependence
    on $M_1$ is very weak.}
\end{figure}

At this moment the only free parameter left is the VEV ratio $x$, and
everything can be expressed via $x$, $M_1$, $M_2$, and $m_2\simeq\Delta
m_\mathrm{sol}$, $m_3\simeq\Delta m_\mathrm{atm}$.  From
Eqs.~\eqref{eqn:typeIIsee-sawmod} and \eqref{eqn:x}, we get
\be\label{eqn:v_R}
v_R=\sqrt{\frac{v^2x}{\frac{m_2}{M_2}+\theta_2^2}}
\;.
\ee
The VEV of the left-handed triplet $\Delta_L$ is then given by
\be
v_L=\sqrt{v^2x\left(\frac{m_2}{M_2}+\theta_2^2\right)}
\;.
\ee
Together with Eq.~\eqref{eqn:v_R}, Eq.~\eqref{eqn:theta_2^2typeII} determines the proportionality constant $p$:
\be
p = \sqrt{\theta_2^2v_R^2/v^2}
\;.
\ee
The full mass matrix \eqref{eqn:see-saw12} is then given by
\begin{equation}
  \mathcal{M} =
  \begin{pmatrix} U^\star & 0 \\ 0 & U^\star \end{pmatrix}
  \left(
    \baz
    M_{L}^\mathrm{diag} & \ m_D^\mathrm{diag} \\
    m_D^\mathrm{diag} & M_{R}^\mathrm{diag}
    \ea
  \right)
  \begin{pmatrix} U^\dagger & 0 \\ 0 & U^\dagger \end{pmatrix}
  \;,
\end{equation}
where
\begin{subequations}
\begin{align}
m_D^\mathrm{diag}&=p \frac{v}{v_L} M_{L}^\mathrm{diag}=p \frac{v}{v_R}M_{R}^\mathrm{diag}
\;,\label{eqn:propor1}\\
M_{L}^\mathrm{diag}&=\frac{v_L}{v_R}M_{R}^\mathrm{diag}
\;,
\end{align}
\end{subequations}
and
\be
M_{R}^\mathrm{diag} = \diag\left(M_1,M_2,M_3\right)
\;.
\ee
Let us fix now the input values.  It was mentioned before that
$x\sim\mathcal{O}(1)$ is natural in the LR symmetric model,
therefore we simply choose $x=1$.  For the masses of the DM and the entropy
producing sterile neutrinos, we take the smallest possible ones (see
Sec.~\ref{sec:cosm-requ}):
\begin{subequations}\label{eqn:M_1M_2}
 \begin{align}
  M_1&=1.6\keV
\;,\\
  M_2&=2.7\GeV \label{eqn:M_2}
\;.
 \end{align}
\end{subequations}
With this input, we obtain
\begin{align}\label{eqn:output}
  m_1 &= 5.2\times10^{-9}\eV\;,\notag\\
  m_2&\simeq\sqrt{\Delta m_\mathrm{sol}^2}=8.7\times10^{-3}\eV\;,\notag\\
  m_3&\simeq\sqrt{\Delta m_\mathrm{atm}^2}=4.9\times10^{-2}\eV
\;,\notag\\
M_3&=15.1\GeV\;,\notag\\
\theta_1^2=\theta_2^2=\theta_3^2&=2.3\times10^{-15}
\;,\notag\\
v_R&=9.67\times10^4\TeV
\;,\notag\\
v_L&=313\keV
\;,\notag\\
p&=0.027
\;.
\end{align}
We also plot the values of $\theta_1^2$ and $v_R$ for several $M_1$
and $M_2\gtrsim2.4\GeV$ in the Figs.~\ref{fig:theta_2-M2} and
\ref{fig:M2-vR}.  Because of the smallness of $\theta_2^2$ compared to
$m_2/M_2$ and its suppression with $M_1^2$ or $M_2^3$ (see
Eq.~\eqref{eqn:theta_2^2}), $v_R$ given by Eq.~\eqref{eqn:v_R} is
effectively independent of $\theta_2^2$.  Therefore, the curve of $v_R$
has only a very weak $M_1$ dependence.
However, for bigger $M_1$ one has to account for
the BBN bound \eqref{BBNbound} on $M_2$.

One can check that none of the bounds, summarized in
Sec.~\ref{sec:summary} is violated.  Also the mixing angle
$\theta_1^2$ corresponding to our DM neutrino is much lower than its
upper bound, Eq.~\eqref{theta-Xraybound}.  However, we should also choose
the Higgs potential to have very small mixing between the left and
right gauge bosons (see Eqs.~(\ref{zetabound}) and \eqref{eqn:tan(2zeta)}).

The right-handed scale $v_R$ is large, and the additional gauge and Higgs
bosons are not observable (they all have masses $\propto v_R$).
The famous $\rho$ parameter 
 \be \rho\equiv\frac{M_W^2}{M_Z^2\,\cos^2\theta_W}
= 1 \;, \ee
which is equal to 1
at tree level in SM also gets a negligible correction which is equal to 
 \be
\rho=\frac{v^2+2\,\abs{v_L}^2}{v^2+4\,\abs{v_L}^2} \;, 
\ee
in the LR symmetric model \cite{Gunion:1989we}.
For the
small $v_L$ of the order of MeV, the deviations are well below
the current experimentally allowed deviation of the order
${\mathcal O}(10^{-4})$ \cite{Amsler:2008zzb}.

%%%%%%%%%%%%%%%%%%%%%%%%%%%%%%%%%%%%%%%%%%%%%%%%%%%%%%%%%%%%%%%%%%%%%%%%
\section{Conclusions}
In this paper, we analyzed the possibility to have a keV scale sterile
neutrino warm dark matter in gauge extensions of the standard model. We found that it is
possible to circumvent the na\"ive expectation of significant
overproduction of dark matter in case of a light particle (sterile
neutrino) decoupling from the thermal equilibrium while still
relativistic.  The possible ways out include a low reheating temperature
(so that the thermal equilibrium is never reached by the would be DM
sterile neutrino), (very) large number of degrees of freedom in the
early universe at the DM neutrino freeze-out, or subsequent dilution of
its density by the out-of-equilibrium decay of a massive particle
(another sterile neutrino).  We further analyze this last possibility
as being the most natural\footnote{Another natural possibility is
  achieved in the $\nu$MSM model \cite{Asaka:2005an,Asaka:2005pn},
  where the sterile neutrinos are the \emph{only} extension of the SM,
  and then the keV sterile neutrino does not enter thermal equilibrium
  up to Planck scale temperatures.}
and formulate a set of requirements for this scenario.
   In
 short, these requirements bound the mass of the DM sterile neutrino
 from below from structure formation considerations, limit its
 mixing angle with active neutrinos and constrain mixing between the SM
 (left) and additional (right) gauge bosons from the radiative decay of
 the DM sterile neutrino, fix the lifetime of the heavier sterile
 neutrino from the requirement of the dilution of the DM abundance down
 to the observed value, and finally constrain the mass of these heavier
 sterile neutrinos from the big bang nucleosynthesis considerations.

We demonstrated in this scenario that the type I low scale seesaw mechanism of
generating masses for the active neutrino can not lead to sufficient
dilution of the DM abundance.  At the same time, we provide a working
example, where the active neutrinos are generated by a type II style
seesaw in the context of an exactly LR symmetric theory.  The
provided general constraints and observations can serve as a basis for
the search of a grand unified theory with WDM sterile neutrinos.

\begin{acknowledgments}

  F.B. thanks M.~Shaposhnikov and S.~Sibiryakov for helpful comments
  and O.~Ruchayskiy for discussion on the Lyman-$\alpha$ bound.
  This work has been supported in part by the DFG Grant No. SFB-TR27 ``Neutrinos and Beyond.''
\end{acknowledgments}

%%%%%%%%%%%%%%%%%%%%%%%%%%%%%%%%%%%%%%%%%%%%%%%%%%%%%%%%%%%%%%%%%%%%%%%%
\appendix

\section{Decay widths of a sterile neutrino}
\label{appx:widths}

In the mass range $2.7\GeV\leq M_2 <M_{W}$, a sterile neutrino $N_2$ dominantly decays into leptons and spectator quarks.
The corresponding partial widths can be calculated in the $\nu$MSM,
because additional boson interactions, which usually appear in more
complicated models, are of high scale---of order $\mathcal{O}(M)$---compared to that of the electroweak scale. To make use of the $\nu$MSM results, we have to compare, for being on the safe side, the suppression factors in the $\nu$MSM  $\sim\mathcal{O}(\abs{\theta_i}^2)$ with that of additional interactions $\sim\mathcal{O}(M_W/M)^4$ which appear in the models we are interested in. Furthermore, the mixing of the new bosons should be small compared to $\mathcal{O}(\abs{\theta_i})$; otherwise, there could be significant contributions from processes where new bosons mix with the SM ones (see, for example, Appendix \ref{sec:raddecay}).
These effects are neglected in the following calculations.  One can
see that for most practical purposes it is the case, as far as the
bound on the gauge boson mixings (\ref{zetabound}) is much stronger
then those for the active-sterile neutrino mixings.

Moreover, these additional contributions do not affect the conclusions
in the main part of the article.
Really, in Sec.~\ref{sec:typeI}
additional interactions can only result in a stronger bound and therefore the conclusion remains the same.
In the type II seesaw model discussed in Sec.~\ref{sec:typeII}, we need the exact value of the width. However, if we assume no mixing of the $W$ bosons ($\zeta=0$), the contributions of the $W_R$ boson mediated processes are in the considered mass range negligible small (cf. Figure \ref{fig:theta_2-M2}).

In the following, we give all relevant formulas for the decay rates (at tree level)  of a sterile neutrino $N_2$ with a mass $M_2$ above the BBN bound $2.7\GeV$ (cf. \eqref{BBNbound}) and below the SM $W$-boson mass $M_{W}\simeq80\GeV$ \cite{Gorbunov:2007ak}:

\begin{widetext}
\begin{subequations}\label{eqn:threebodylepton}
{%\footnotesize
\begin{align}
\Gamma_1\left( N_2\to \sum_{\alpha,\beta}\nu_\alpha\bar\nu_\beta \nu_\beta
\right)&=
\frac{G_F^2M_2^5}{192\pi^3}\cdot\sum_\alpha |\theta_{2\alpha}|^2\;,\\
\Gamma_2\left( N_2\to l^-_{\alpha\neq\beta}l^+_\beta  \nu_\beta \right)&=
\frac{G_F^2M_2^5}{192\pi^3}\cdot |\theta_{2\alpha}|^2 \left(
1-8x_l^2+8x_l^6-x_l^8-12x_l^4 \log x_l^2 \right)\;,~~~x_l=\frac{{\rm max }\left[
  M_{l_\alpha},\;M_{l_\beta}\right]}{M_2}\;,  \\
\Gamma_3\left( N_2\to \nu_\alpha l_\beta^+l_\beta^- \right)&=
\frac{G_F^2M_2^5}{192\pi^3}\cdot |\theta_{2\alpha}|^2 \cdot
\Biggl[\left(
  C_1\cdot(1-\delta_{\alpha\beta})+C_3\cdot\delta_{\alpha\beta}\right)
\biggl(
\left( 1-14x_l^2-2x_l^4-12x_l^6\right)\sqrt{1-4x_l^2}\nonumber\\&
+12 x_l^4 \left( x_l^4-1\right) L \biggr)
+ 4
\left( C_2\cdot(1-\delta_{\alpha\beta}) +C_4\cdot\delta_{\alpha\beta}  \right)
\biggl( x_l^2 \left( 2+10 x_l^2-12 x_l^4\right) \sqrt{1-4x_l^2} \nonumber\\
&+6x_l^4\left( 1-2 x_l^2+2x_l^4\right) L\biggl)\Biggr]\;,
\end{align}
}
with
{%\footnotesize
\[
L=\log\left[
\frac{1-3x_l^2-\left( 1-x_l^2 \right)\sqrt{1-4x_l^2}}
{x_l^2\left( 1+\sqrt{1-4 x_l^2}\right)}\right]\;,~~~~x_l\equiv\frac{M_l}{M_2}\;,
\]
}
and
{%\footnotesize
\begin{align*}
C_1&=\frac{1}{4}\left( 1-4\sin^2\theta_w+8\sin^4\theta_w\right)\;,
&C_2=\frac{1}{2}\sin^2\theta_w\left( 2\sin^2\theta_w-1\right)\;,\\
C_3&=\frac{1}{4}\left( 1+4\sin^2\theta_w+8\sin^4\theta_w\right)\;,
&C_4=\frac{1}{2}\sin^2\theta_w\left( 2\sin^2\theta_w+1\right)\;.
\end{align*}
}

\end{subequations}
\end{widetext}

\begin{figure}
 \includegraphics{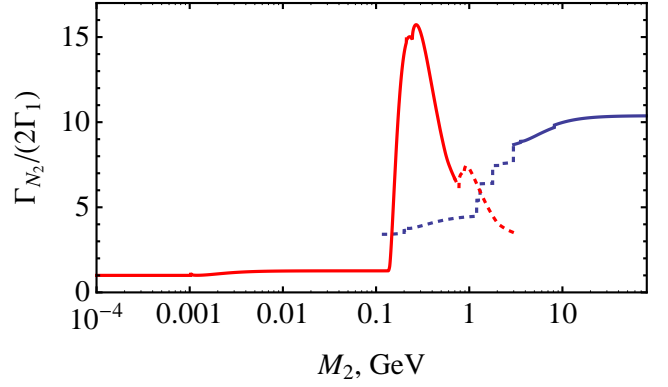}
 \caption{\label{fig:GammaRatios} The ratio $\Gamma_{N_2}/2\Gamma_1$
   calculated in the specific LR model of Sec.~\ref{sec:typeII}. We
   used Eq.~\eqref{eqn:threebodylepton} together with the formulas
   for the two-body decays into mesons \cite{Gorbunov:2007ak} for the
   left (red) curve and Eq.~\eqref{eqn:threebodylepton} together with
   Eq.~\eqref{eqn:theebodyquark} for the right (blue) curve. The dotted
   curves correspond to the region, where both approximations are not
   entirely reliable.}
\end{figure}

The formulas for the decay modes into quarks are presented below.
In the range $2.7\GeV\leq M_2<M_W$, it is sufficient to use the free
quark approximation for the decay products.
We give these formulas in the approximation where $M_2$ is much
heavier than the decay product masses (unlike above
(Eqs.~\eqref{eqn:threebodylepton}) for the lepton decays).
The corrections are important at the threshold, when new decay
channels open.  However, at high mass $M_2$, this introduces a rather
small relative error, because the number of open channels into light
particles is significant and provides the main part of the decay
width.  The exact analysis would
smooth the discontinuities of the decay width at the mass thresholds 
(Fig.~\ref{fig:GammaRatios}):
\begin{subequations}\label{eqn:theebodyquark}
{%\footnotesize
\begin{align}
\Gamma_4\left( N_2\to l^{-}_\alpha U \bar{D} \right)&=
\frac{G_F^2M_2^5}{192\pi^3}\cdot3\cdot |V_{UD}|^2\cdot|\theta_{2\alpha}|^2\;,
\\
\Gamma_5\left( N_2\to \nu_\alpha q \bar{q} \right)&=
\frac{G_F^2M_2^5}{192\pi^3}\cdot3\cdot \varXi^q  \cdot |\theta_{2\alpha}|^2\;,
\end{align}
}
with
{%\footnotesize
\begin{align*}
\varXi^q = (g_L^\nu)^2\cdot\left((g_L^q)^2+(g_R^q)^2\right)
\;.
\end{align*}
}
\end{subequations}
The factor $3$ is the color factor and $V_{UD}$ are the Cabibbo-Kobayashi-Maskawa--matrix elements.
The coupling constants $g_L$ and $g_R$ correspond to the coupling of the
$Z$ boson to left- or right-handed particles, respectively. For a fermion $f$ with weak isospin component
$I_3^f$ and charge $q_f$, one has
\begin{subequations}
{%\footnotesize
\begin{align}
g_L^f&=I_3^f-q_f\,\sin^2{\theta_W}\;, \\
g_R^f&=-q_f\,\sin^2{\theta_W}\;.
\end{align}
}
\end{subequations}
Table~\ref{tab:couplingConst} gives the required values of the charges.
\begin{table}
\begin{ruledtabular}
 \begin{tabular}{l|l|l}
fermions & $g_L^q$ & $g_R^q$ \\\hline
$\nu_e, \nu_\mu, \nu_\tau$	&$g_L^\nu=\frac{1}{2}$	&$g_R^\nu=0$\\
$U = u,c,t$	
&$g_L^U=\frac{1}{2}-\frac{2}{3}s_W^2$&$g_R^U=-\frac{2}{3}s_W^2$\\
$D = d,s,b$	
&$g_L^D=-\frac{1}{2}+\frac{1}{3}s_W^2$&$g_R^D=\frac{1}{3}s_W^2$\\
 \end{tabular}
\end{ruledtabular}
\caption{\label{tab:couplingConst}Coupling constants}
\end{table}
In the mass range of $M_2$ mentioned above, the Majorana neutrino total decay rate $\Gamma_{N_2}$ is a sum of all rates presented
above multiplied by a factor of 2, which accounts for charge-conjugated
decay modes.

When $M_2$ exceeds the SM $Z$-boson mass $M_Z=91\GeV$ and if contributions of new interactions are negligible, the sterile neutrino predominantly decays into a SM gauge boson and a lepton. Then its total decay width is given by 
{%\footnotesize
\begin{align}\label{eqn:N2decayWZ}
 \Gamma_{N_2}&=2\frac{G_F M_2^3}{8\sqrt{2}\pi}\left[\left(1+\frac{2 M_W^2}{M_2^2}\right)\left(1-\frac{M_W^2}{M_2^2}\right)^2\right.\notag\\
&\qquad\left.+\frac{1}{2}\left(1+\frac{2M_Z^2}{M_2^2}\right)\left(1-\frac{M_Z^2}{M_2^2}\right)^2\right]\cdot \sum_\alpha |\theta_{2\alpha}|^2
\;.
\end{align}
}
In between, where $M_W\leq M_2<M_Z$, one can approximate the width by the first term in Eq.~\eqref{eqn:N2decayWZ}.

Below $M_2\sim2\GeV$, it is important to consider mesons instead of
quarks as final states for the sterile neutrino decay.  Therefore,
instead of the three-body decay modes into spectator quarks
\eqref{eqn:theebodyquark}, one has to use the corresponding two-body
ones into mesons; see \cite{Gorbunov:2007ak} for the decay width
formulas.  In our case, this mass range of $M_2$ is forbidden by the
BBN bound \eqref{BBNbound} and therefore we do not list them here.
Nevertheless, to get a feeling of the behavior of the total width
$\Gamma_{N_2}$, 
we show in Fig.~\ref{fig:GammaRatios} the ratio
$\Gamma_{N_2}/2\Gamma_1$ calculated in the specific model described in
Sec.~\ref{sec:typeII}, using both the free quark and chiral meson
approximations.  It is clearly seen that the transition between two
approximations happens around 1~GeV.
Because of Eq.~\eqref{eqn:thetaSecTypeII}, the total
decay width is proportional to $\theta_2^2$ and therefore the plotted
ratio is independent of this quantity.  In more general models,
Eq.~\eqref{eqn:thetaSecTypeII} will no longer be valid. However, as one
recognizes by considering the formulas together with the definition of
$\theta_2$ (cf. Equation \eqref{eq:theta2def}), there will be no
significant difference, especially for heavy masses $M_2$, so that
Fig.~\ref{fig:GammaRatios} can be used as a good estimate in such
models. Note that in the region $M_2\sim{\mathcal O}(1)\GeV$ (dotted
in Fig.~\ref{fig:GammaRatios}), decays into spectator quarks more and
more replace decays into mesons and therefore one has to carefully
reanalyze the given formulas, if one is interested in this mass range.

%%%%%%%%%%%%%%%%%%%%%%%%%%%%%%%%%%%%%%%%%%%%%%%%%%%%%%%%%%%%%%%%%%%%%%%%
\section{Radiative decay width}
\label{sec:raddecay}

Here, we give some details of calculation of the width for the radiative decay $N_1\to\gamma\nu_i$ shown in Fig.~\ref{fig:DecayN}.
We will follow Ref.~\cite{Lavoura:2003xp}, where general formulas for this type of process are given. In our case, $N_1$ denotes a heavy sterile neutrino with mass $M_1$, $\nu_i$ one of the active neutrinos with mass $m_i$, and $\gamma$ a photon. The neutrinos are considered as Majorana particles.

The amplitude for such a decay is $e\epsilon^\star_\mu(q){\mathcal M}^\mu$, where $e$ is the electric charge of the positron and $\epsilon^\star_\mu(q)$ the polarization vector of the outgoing photon. The Ward identity for the electromagnetic current implies that $q_\mu {\mathcal M}^\mu$ must be zero; therefore ${\mathcal M}^\mu$ must have the form
\begin{equation}
 {\mathcal M}^\mu = \bar{u}_i\left[i\sigma^{\mu\nu}q_\nu(\sigma_L L+\sigma_R R)\right]u_1
\;,
\end{equation}
where $\sigma^{\mu\nu}=(i/2)[\gamma^\mu,\gamma^\nu]$, $L = (1-\gamma_5)/2$ and $R = (1+\gamma_5)/2$ are the projectors of chirality. $\sigma_L$ and $\sigma_R$ are numerical coefficients with dimension of inverse mass. 
The partial decay width for $N_1\to\nu_i\gamma$ is then given by
\begin{equation}\label{eqn:partialdecay}
 \Gamma_{N_1\to\gamma\nu_i}=\frac{(M_1^2-m_i^2)^3}{16\pi M_1^3}\left(\abs{\sigma_L}^2+\abs{\sigma_R}^2\right)
\;.
\end{equation}
By comparing the Lagrange term for the charged current \eqref{eq:Lcc} combined with Eq.~\eqref{eq:Wmix} and the transformation rule which diagonalizes the neutrino mass matrix in Eq.~(\ref{eq:Lmass})
\begin{equation}\label{eqn:trafoexact}
  \begin{pmatrix}
    \tilde\nu_{aL}\\
    \tilde N_{aR}^c
  \end{pmatrix}
  =
  \left(\begin{array}{cc}
    A & B \\
    C & D
   \end{array}\right)
  \begin{pmatrix}
    \nu_{iL}\\
    N_{IR}^c
  \end{pmatrix}
  \;,
\end{equation}
with that given in Chapter 5 of \cite{Lavoura:2003xp}, we can easily calculate the coefficients $\sigma_L$ and $\sigma_R$.
Supposing from the very beginning, that the right-handed scale is much larger than the left one, $M\gg M_{W}\simeq80.4\GeV$, and neglecting the active neutrino masses,
we get\footnote{Note that our results do not coincide with the formulas in \cite{Chattopadhyay:1986cj}. This is because of a mistake in the second term of the third line of Ew. (10) in \cite{Chattopadhyay:1986cj}. The correct labeling of the transformation matrices should be $P_{aB}Q_{aA}$ instead of $P_{aA}Q_{aB}$. In our notations, where a sterile neutrino (with mass eigenstate index 1) decays through the radiative process into an active neutrino (with mass eigenstate index $i$), the expression $P_{aB}Q_{aA}$ translates into $B_{a1}C_{ai}$ which is contained in Eq.~\eqref{eqn:sigmaL}.}
\begin{subequations}\label{eqn:sigmas}
\begin{align}
i\sigma_R=&\frac{g^2\,e}{32\,M_{W}^2\,\pi^2}\times\notag\\
&\sum_{a=e,\mu,\tau}\big\{\cos\zeta\sin\zeta\,A^\star_{ai}D^\star_{a1}\,m_{l_a} {\mathcal F}(r_a) \notag\\
&\qquad\qquad+\cos^2\zeta\,A^\star_{ai}B_{a1}M_1 F(r_a)\big\}\;,\label{eqn:sigmaR}\\
i\sigma_L=&\frac{g^2\,e}{32\,M_{W}^2\,\pi^2}\times\notag\\
&\sum_{a=e,\mu,\tau}\big\{\cos\zeta\sin\zeta\,C_{ai}B_{a1}\,m_{l_a} {\mathcal F}(r_a)\big\}\label{eqn:sigmaL} \;,
\end{align}
\end{subequations}
where $F(r_a)$ and ${\mathcal F}(r_a)$ are functions of $r_a\equiv m_{l_a}^2/M_{W}^2$. In our case, we have in good approximation $F(r_a)\simeq-3/2$ and ${\mathcal F}(r_a)\simeq4$. The exact expressions for these functions were calculated by us and do agree with that given in Ref.~\cite{Chattopadhyay:1986cj}.

Because of the Majorana nature of our ingoing and outgoing neutrinos, we also have to add the contribution of the complex conjugated process to our amplitude. This is easily obtained out of Eq.~\eqref{eqn:sigmas} by putting in the substitutions
\begin{equation}
 \begin{array}{ccc}
  A,B \to A^\star,B^\star &\;\text{and}\;& C,D \to C^\star,D^\star\;,\\
  \gamma_5\to-\gamma_5 &\Rightarrow& L,R \to R,L\;,
 \end{array}
\end{equation}
and an overall negative sign coming from the photon vertex.
After adding the derived $\sigma_L$ and $\sigma_R$, it is easy to see that $\abs{\sigma_L}^2 = \abs{\sigma_R}^2$, where
\begin{align}
 \abs{\sigma_L}^2=&\left(\frac{g^2\,e}{32\,M_{W}^2\,\pi^2}\right)^2\times\notag\\
&\qquad\Bigg|4\cos\zeta\sin\zeta\,\sum_{a=e,\mu,\tau}(A_{ai}D_{a1}-C_{ai}B_{a1})\,m_{l_a}\notag\\
&\qquad\qquad\qquad-\frac{3}{2}\cos^2\zeta\left(\sum_{a=e,\mu,\tau}\,A_{ai}B_{a1}^\star\right)\, M_1\Bigg|^2\;.
\end{align}
By putting this into Eq.~\eqref{eqn:partialdecay}, we obtain
\begin{align}\label{eqn:DecayWidth}
  \Gamma_{N_1\rightarrow\gamma\nu_i} &\simeq
    \frac{G_F^2\,\alpha\, M_1^3}{64\,\pi^4}\times \notag\\
&\Bigg|4\,\cos \zeta \, \sin
\zeta \,\sum_{a\,=\,e,\mu,\tau} (A_{a i}\, D_{a 1}- C_{a i}\,B_{a 1})\,m_{l_a}\notag\\
&\qquad\qquad-\frac 32 \cos^2 \zeta \left(\sum_{a\,=\,e,\mu,\tau} \,
A_{ai}\,B_{a1}^\star\right)\,M_1\Bigg|^2
.
\end{align}
Here, $G_F$ is the Fermi constant, $\alpha$ is the fine-structure constant, and $m_{l_a}$ is the mass of the charged lepton propagating in the loop.

The total width of the radiative decay is given by 
\begin{equation}
\Gamma_{N_1\rightarrow\gamma\nu}= \sum_{i=1}^3\Gamma_{N_1\rightarrow\gamma\nu_i}
\;.
\end{equation}
In a model where a seesaw mechanism of type I or type II is responsible for the small active neutrino masses, the transformation \eqref{eqn:trafoexact} is given by Eq.~\eqref{eqn:trafo}. Putting this into our formulas, we get out of Eq.~\eqref{eqn:DecayWidth} the expression \eqref{eqn:zetanonzero}.

%%%%%%%%%%%%%%%%%%%%%%%%%%%%%%%%%%%%%%%%%%%%%%%%%%%%%%%%%%%%%%%%%%%%%%%%
\section{Casas-Ibarra Parametrization}
\label{CasasIbarra}
In this part of the Appendix, we describe the approach of parametrizing the
Dirac-Yukawa matrix, which was proposed by Casas and Ibarra
\cite{Casas:2001sr}. Here, we want to give a short review of the generalised version which also applies to the type II seesaw mechanism \cite{Akhmedov:2008tb}.

Let us consider a Majorana mass matrix with the pattern
\be
\left(
\baz
M_L & m_D \\
 m_D^T & M_R
\ea
\right)=\left(
\baz
f_L\,v_L & y\,v \\
 y^T\,v & f_R \,v_R
\ea
\right)
\;.
\ee
The type II seesaw formula can be written in the form
\be \label{eqn:m_nu}
m_\nu - M_L = -m_D M_R^{-1} m_D^T
\;,
\ee
where $m_\nu$ is the active neutrino mass matrix.
Let us define the symmetric and in general complex $3\times3$ matrix
\be
X_\nu \equiv m_\nu - M_L
\;.
\ee
This matrix and $M_R$ can be diagonalized by unitary transformations:
\begin{subequations}\label{eqn:XMRtrafo}
\begin{align}
X_\nu &= V_\nu^\star \, X_\nu^{\diag} \,V_\nu^\dagger =\left[ V_\nu^\star(X_\nu^{\diag})^{\frac
12}\right]\left[V_\nu^\star(X_\nu^{\diag})^{\frac 12}\right]^T\;, \label{eqn:X}\\
M_R&=V_R^\star M_R^{\diag}V_R^\dagger\;. \label{eqn:M_Rtrafo}
\end{align}
\end{subequations}
Multiplying Eq.~\eqref{eqn:m_nu} by $\left[ V_\nu^\star(X_\nu^{\diag})^{\frac
12}\right]^{-1}$ from the left and by $\left\{\left[V_\nu^\star(X_\nu^{\diag})^{\frac
12}\right]^T\right\}^{-1}$ from the right and using Eq.~\eqref{eqn:XMRtrafo}, we find
\be\label{eqn:IRRT}
I = R \, R^T 
\;,
\ee
with
\be
 R = \pm i \left(X_\nu^{\diag}\right)^{- \frac 12}V_\nu^T m_D V_R\left(M_R^{\diag}\right)^{-\frac 12 }
\;.
\label{eqn:RRT}
\ee
Equations \eqref{eqn:IRRT} and \eqref{eqn:RRT} mean that the type II seesaw relation requires $R$ to
be a complex orthogonal matrix, but otherwise does not constrain it. In this way
we obtain for the Dirac-type Yukawa coupling in the basis where $M_R$ is diagonal
\be
m_D=v y = \pm i\, V_\nu^\star\sqrt{X_\nu^{\diag}}\,R\,\sqrt{M_R^{\diag}}V_R^\dagger
\label{eqn:casas1}
\;,
\ee
where R is an arbitrary complex orthogonal matrix. It can be parametrized as
\be \label{eqn:R}
R= \pm\,R_{12}\,R_{13}\,R_{23}
\;,
\ee
where $R_{ij}$ is the matrix of rotation by a complex angle $\omega_{ij}$ in the $ij$ plane.
This is the so-called Casas-Ibarra parametrization of the Dirac-Yukawa \cite{Casas:2001sr}. Note
that this parametrization has its origin in the difference of the number of
high energy and low energy parameters. There are less low energy parameters,
because the high energy ones are integrated out.
The latter cannot influence the low energy theory, and therefore can be
parametrized arbitrarily.

The formula for the type I seesaw can easily derived out of \eqref{eqn:casas1}. Because of $M_L =0$, $X_\nu$ corresponds in this case to the active neutrino mass matrix $m_\nu$. Thus the basis transformation matrix $V_\nu$ in Eq.~\eqref{eqn:X} is the PMNS matrix $U$ and we arrive at 
\be \label{eqn:CasasIbarra}
m_D = vy= \pm i\,U^\star\sqrt{m_\nu^{\diag}}\,R\,\sqrt{M_R^{\diag}}V_R^\dagger
\;.
\ee
In chapter~\ref{sec:typeI} we make use of this parametrization.

%%%%%%%%%%%%%%%%%%%%%%%%%%%%%%%%%%%%%%%%%%%%%%%%%%%%%%%%%%%%%%%%%%%%%%%%
\bibliographystyle{h-physrev-4new2}
\bibliography{all}

\end{document}